\documentclass[structabstract]{aa}  
\usepackage{graphicx}
\usepackage{txfonts}
\usepackage{multirow}
\usepackage{longtable,lscape}
\usepackage[]{natbib}
\bibpunct{(}{)}{;}{a}{}{,}
\usepackage{stfloats}
\usepackage{fixltx2e}
\usepackage{textcomp}
\begin{document}
    \title{The dark nature of GRB\,130528A and its host galaxy
}


\author{S. Jeong \inst{1}
\and
A. J. Castro-Tirado \inst{1,2}
\and
M. Bremer \inst{3}
\and
J. M. Winters \inst{3}
\and
J. Gorosabel \inst{1,4,5}
\and
S. Guziy \inst{6}
\and
S. B. Pandey \inst{7}
\and
M. Jel\'inek \inst{1}
\and
R. S\'anchez-Ram\'irez \inst{1}
\and
Ilya V. Sokolov \inst{8,9}
\and
N. V. Orekhova \inst{10,11}
\and
A. S. Moskvitin \inst{12}
\and
J. C. Tello \inst{1}
\and
R. Cunniffe \inst{1}
\and
O. Lara-Gil \inst{1} 
\and
S. R. Oates\inst{1}
\and
D. P\'erez-Ramírez \inst{13}
\and
J. Bai \inst{14,15}
\and
Y. Fan \inst{14,15,16}
\and
C. Wang \inst{14,15,16}
\and
I. H. Park \inst{17}
}

\institute{
Instituto de Astrof\' isica de Andaluc\'ia (IAA-CSIC), Glorieta de la Astronom\'ia s/n, 
E-18008, Granada, Spain.\\
\email{sjeong@iaa.es}
\and
Unidad Asociada Departamento de Ingeniería de Sistemas y Automática, E.T.S. 
de Ingenieros Industriales, Universidad de Málaga, Spain.
\and 
Institute de Radioastronomie Millim\'etrique (IRAM), 300 rue de la Piscine, 38406 
Saint Martin d' H\'eres, France.
\and
Unidad Asociada Grupo Ciencias Planetarias UPV/EHU-IAA/CSIC, Departamento 
de Física Aplicada I, E.T.S., Ingeniería, Universidad del Pa\'is Vasco UPV/EHU, Bilbao, Spain.
\and
Ikerbasque, Basque Foundation for Science, Bilbao, Spain.
\and
Nikolaev National University, Nikolska 24, Nikolaev 54030, Ukraine.
\and
Aryabhatta Research Institute of Observational Sciences, Manora Peak, Nainital - 263 002, India.
\and
Terskol Branch of Institute of Astronomy Russian Academy of Sciences, Elbrus ave., 81/33, Tyrnyauz, Kabardino-Balkaria Republic, 361623, Russia.
\and
Saint Petersburg State University, Universitetskiy ave 28, Petrodvorets, Saint Petersburg, 198504, Russia.
\and
Petrozavodsk State University, Lenin Str., 33, 185910, Petrozavodsk, Russia.
\and
Open joint stock company, "Scientific Production Corporation System of Precision Instrument Making", Aviamotornaya, 53, Moscow, Russia.
\and
Special Astrophysical Observatory, Nizhniy Arkhyz, Zelenchukskiy region, Karachai-Cherkessian Respublic, 369167, Russia.
\and
Universidad de Ja\' en, Campus Las Lagunillas, s/n, Ja\'en, Spain.
\and
Yunnan Observatories, Chinese Academy of Sciences, Kunming 650011, China.
\and
Key Laboratory for the Structure and Evolution of Celestial Objects, Chinese Academy of Sciences, Kunming 650011, China.
\and
University of Chinese Academy of Sciences, Beijing 100049, China.
\and
Department of Physics, Sungkyunkwan University, Suwon, Korea. 
}

\date{Received; accepted Aug 20, 2014}

\abstract
{}
{We study the dark nature of GRB\,130528A through multi-wavelength observations and conclude that the main reason for the optical darkness is local extinction inside of the host galaxy.}
{Automatic observations were performed at the Burst Optical Observer and Transient Exploring System (BOOTES)-4/MET robotic telescope. 
We also triggered target of opportunity (ToO) observations at Observatorio de Sierra Nevada (OSN), IRAM Plateau de Bure Interferometer (PdBI) 
and Gran Telescopio Canarias (GTC + OSIRIS). The host galaxy photometric observations in optical to near-infrared 
(nIR) wavelengths were achieved through large ground-based aperture telescopes, such as 10.4m Gran Telescopio Canarias (GTC), 
4.2m William Herschel Telescope (WHT), 6m Bolshoi Teleskop Alt-azimutalnyi (BTA) telescope, and 2m Liverpool Telescope (LT). 
Based on these observations, spectral energy distributions (SED) for the host galaxy and afterglow were constructed.}
{Thanks to millimetre (mm) observations at PdBI, we confirm the presence of a mm source within the XRT error circle that faded over the course of our observations and identify 
the host galaxy. However, we do not find any credible optical source within early observations with BOOTES-4/MET and 1.5m OSN telescopes. 
Spectroscopic observation of this galaxy by GTC showed a single faint emission line that likely corresponds to [OII] 3727$\AA$ at a 
redshift of 1.250 $\pm$ 0.001, implying a star formation rate (SFR)($M_{\odot}$/yr) $\textgreater$ 6.18 $M_{\odot}$/yr without correcting for dust extinction. 
The probable line-of-sight extinction towards GRB\,130528A is revealed through analysis of the afterglow SED, resulting in a value of 
$A^{GRB}_{V}$ $\geq$ 0.9 at the rest frame; this is comparable to extinction levels found among other dark GRBs. 
The SED of the host galaxy is explained well ($\chi^{2}$/$d.o.f.$=0.564)
by a luminous (M$_{B}$=-21.16), low-extinction ($A_{V}$=0, rest
frame), and aged (2.6 Gyr) stellar population. We can explain this
apparent contradiction in global and line-of-sight extinction 
if the GRB birth place happened to lie in a local dense environment. In light of having relatively small specific
SFR (SSFR) $\sim$ 5.3 $M_{\odot}$/yr $(L/L^{\star})^{-1}$, this also could
explain the age of the old stellar population of host galaxy.} {}

\authorrunning{S. Jeong et al.}
\titlerunning{The dark nature of GRB\,130528A and its host galaxy}
\maketitle
%

\section{Introduction}
\indent
Since the launch of the {\it Swift}, $\sim$ 78$\%$ (667/856 as of Apr 1, 2014) of
observed Gamma-ray Bursts (GRBs)\footnote{http://swift.gsfc.nasa.gov/archive/grb$\textunderscore$table/}, were detected accurately and rapidly by the
{\it Swift} X-ray telescope (XRT). Among them 
$\sim$ 73$\%$ (488/667) of GRBs were detected by the {\it Swift}
UV/optical telescope (UVOT) or ground-based telescopes at UV/optical/IR
wavelengths, but UV/optical/IR emission was not detected in 20-27\% of observed GRBs (see also \citealt{Melandri:2012aa,
 Greiner:2011aa}), despite deep searches during several hours by
ground facilities. Events lacking UV/optical/IR emission are dubbed "dark GRBs" \citep{Groot:1998aa} with GRB 970111 being the first such case
\citep{Castro-Tirado:1997aa, Gorosabel:1998aa}.

Currently dark GRBs are defined as those events having no 
UV/optical afterglow but also a relatively low optical-to-X-ray flux ratio 
(see \citealt{Jakobsson:2004aa} and \citealt{Horst:2009}). 
Plausible causes for dark GRBs, such as observational bias, high level of extinction 
within the galaxy, Lyman-$\alpha$ cut-off (for high redshift bursts) and 
intrinsically low UV/optical fluxes are claimed, although a combination of 
two or three causes is likely (also discussed by \citealt{Rol:2005aa, Fynbo:2001aa}). 
The number of well-observed dark GRBs and their hosts is continually being increased thanks to well-targeted ground-based ToO campaigns, 
which enable us to gain better insight into the nature of GRBs and their environments. Moreover, future space-based missions, 
Ultra-Fast Flash Observatory (UFFO)-pathfinder/{\it Lomonosov}, and UFFO might be helpful to understand the dark nature of GRBs 
using the early optical follow-up within several seconds after GRB onset \citep{Park:2013aa, Jeong:2013aa}.

Recent studies have shown that dust extinction inside of the host galaxy might be 
the probable cause of darkness; the GRB is generated in a denser environment compared with optically bright events 
(\citealt{DePasquale:2003aa, Perley:2009aa, Melandri:2012aa}).
The prompt properties of dark GRBs at rest frame do not differ with optically bright events, but interestingly, the average X-ray luminosity 
(unabsorbed X-ray flux at rest frame) of dark bursts is slightly higher, although the observed optical flux is slightly lower (see Fig. 4 in \citealt{Melandri:2012aa}). 
A significant correlation between intrinsic X-ray column density and $\beta_{OX}$ 
has been pointed out by \citet{Campana:2012aa}, and dark GRBs ($\beta_{OX}\textless0.5$, \citealt{Jakobsson:2004aa}) have been shown to 
have a moderately high column density in comparison to optically bright events. 

Host galaxies that harbour dark GRBs are interesting as a study for unbiased samples of star-bursts galaxies in the universe related with SFR \citep{Christensen:2004aa}.
Some hosts of dark GRBs trace a sub-population of massive star-burst galaxies, 
which differ from the main GRB host galaxy population \citep{Rossi:2012aa}. 
\citet{Kruhler:2011aa} report similar results, in that the hosts of the dustiest afterglows have diverse 
properties but are on average redder, more luminous, and massive in comparison to hosts of optically bright events (see also \citealt{Hunt:2014aa}). 
\citet{Perley:2013aa} also deduced similar results 
by investigating 23 dust obscured {\it Swift} GRBs. It suggests that their hosts are more massive, about an order of magnitude, 
compared with unobscured GRBs at similar redshifts.

On May 28, 2013, at 16:41:23 UT, the {\it Swift} Burst Alert Telescope (BAT) 
triggered and located the "North pole'' GRB \citep{DElia:2013aa, Goad:2013aa}. 
The BAT light curve is multiple-peaked with a duration of about 84 s and 
exhibited a peak count rate of $\sim$ 5500 counts/s in the 15-350 keV range at 
$\sim$ 8 s after the trigger. The time-averaged spectrum from T$_{0}$+0.12 
to 79.34 s was fitted by a power law with an exponential cutoff with 
a photon index 1.39 $\pm$ 0.19, E$_{cutoff}$ of 118.3 $\pm$ 79.7 keV and 
total fluence of 5.1 $\pm$ 0.2 $\times$ $10^{-6}$ erg/cm$^{2}$
in the 15-150 keV band \citep{DElia:2013aa, Cummings:2013aa}. 

The {\it Swift}/XRT began observing the field at 64.9 s
after the BAT trigger and found a bright, fading uncatalogued X-ray source \citep{DElia:2013aa}. 
An astrometrically corrected X-ray position was reported later, 
RA(J2000)=$09^{h}$$18^{m}$$0.12^{s}$ and Dec(J2000)=+87\textdegree18'03.7" with an uncertainty of 1.8 arcsec (radius, 90\% confidence, \citealt{Goad:2013aa}).
Initial XRT spectral analysis resulted in a column density of 3.6 $\pm$ 0.6 $\times$ $10^{21}$ cm$^{-2}$ (90 $\%$ confidence, \citealt{Melandri:2013aa}) 
in excess of the galactic value at 8.5 $\sigma$ (5.2 $\times$ $10^{20}$ cm$^{-2}$, \citealt{Kalberla:2005aa}), which shows a high equivalent hydrogen column density $N_{H}$ resembling 
GRB\,051022 \citep{Castro-Tirado:2007aa}. The {\it Swift}/UVOT started follow-up observations 75 s after the BAT 
trigger, however, it did not detect any credible 
afterglow candidate within the XRT error circle down to 
21.7 mag in the white filter \citep{DElia:2013aa, Pasquale:2013aa}. This encouraged ground-based 
observations at several different wavelengths. The 0.4m telescope at ISON-Kitab 
Observatory commenced observations 20 min after the BAT trigger with no 
optical counterpart being reported at a 3 $\sigma$ limit $\textgreater$ 19.1 mag (unfiltered images of 30 s exposure, see \citealt{Volnova:2013aa}).\

In this paper, we discuss the reason for the dark nature of 
GRB\,130528A, using our dataset from the optical 
to mm wavelengths. The structure of our paper is as follows: 
in Section 2, we describe our multi-wavelength observations and data reduction. 
In Section 3, we discuss the observational results, which lead us to consider 
GRB\,130528A as a dark GRB, as well as the host galaxy properties, before summarising
 our conclusions in Section 4.\

\section{Observations and data reduction}
\label{Observations}

\subsection{Photometric observations in optical/near-infrared wavelength}
\label{OptObs}
\indent Following the detection by {\it Swift}/BAT and XRT and the non-detection by 
UVOT, an autonomous search by BOOTES-4/MET was performed prior to a follow-up 
program with several ground-based telescopes. Early time optical observations were carried out at the BOOTES-4/MET 
robotic telescope in Lijiang, China \citep{Castro-Tirado:2012aa}, which automatically responded to the GRB alert 
with observations being conducted on May 28, 
17:52:42 UT, which are $\sim$ 1.1 h after the {\it Swift}/BAT trigger 
(T$_{0}$=16:41:23 UT) in the clear and r-band filters (120 s and 180 s 
exposures, respectively). The field calibration was achieved using GTC deep observations (see below). 
We also triggered the 1.5m OSN located in the Sierra Nevada mountain range in the $I$-band (600 s exposure), 
and the field calibration was conducted using USNO B1.0 catalogue.

Broadband observations in the optical and nIR for the potential 
host galaxy were conducted to produce a SED. 
Observations were performed using various large aperture ground-based telescopes, such as the GTC in r- and i-filters that ranges from $\sim$ 3.4 d 
to $\sim$ 312 d after the burst. The z-band host galaxy observations
were imaged with the 2.0m LT equipped with the
the IO:O instrument during three consecutive nights starting on Apr 4, 
2014. The z-band field calibration was carried out using the 
transformation formulas given by \citet{Jordi:2006aa} and the GTC
r\&i-band secondary standards. A B-band observation was conducted with the 6m BTA telescope with SCORPIO 
located in the Zelenchuksky District on the north side of the Caucasus 
Mountains in southern Russia. The nIR observations in $J$ and $K_{S}$ 
pass-bands using director discretionary time (DDT) were carried out at 
4.2m WHT equipped with LIRIS on Dec 23, 
2013 with seeing $\sim$ 0.8-1.2". The field was calibrated using 
faint standards, FS130 and FS131. The images were reduced, accounting for flat-field and sky background, using the standard method 
within the IRAF. The bias subtraction was performed automatically at the time of data 
saving. The log of optical/nIR observations are 
summarised in Table \ref{table:Optlog}. To determine the photometric magnitudes, we used aperture photometry with 
the DAOPHOT routine in IRAF\footnote{http://iraf.noao.edu}.

\begin{figure}
\centering
\includegraphics[width=8cm]{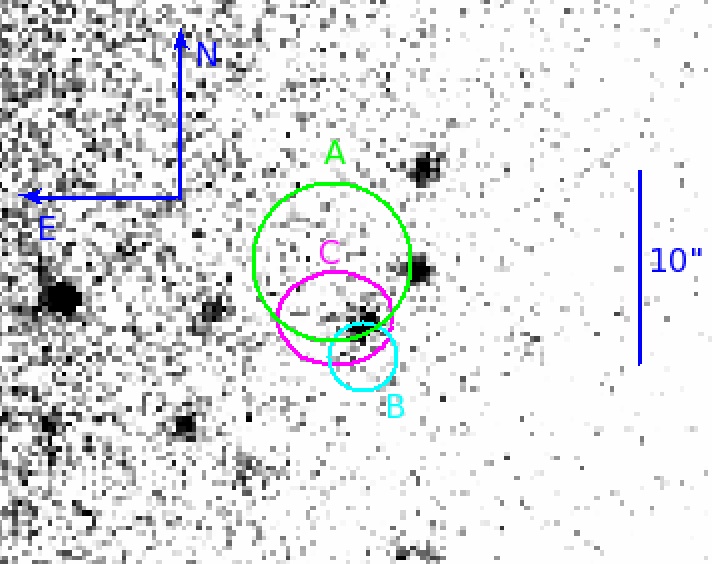}
\caption{The Sloan i-band median combined image (T$_{exp}$=9 $\times$ 100 s) of the field of GRB\,130528A 
taken with the 10.4m GTC on Jun 1, 2013. Circle A \& B represents the XRT error circles 
in 4.1 arcsec \citep{DElia:2013aa} and 1.8 arcsec \citep{Goad:2013aa} radius, respectively. 
The ellipse C shows the mm detection beam size 
by PdBI (see Sect. \ref{MilObs} in this paper). It clearly points out the putative host galaxy of GRB\,130528A by confirming 
a mm afterglow with a signal-to-noise ratio of $\sim$ 4 (at 86.7 GHz). The measured i-band 
magnitude is 22.87 $\pm$ 0.08 in AB system.}
\label{fig:fc}
\end{figure} 

\begin{table*}
\caption{Photometric observations at the GRB\,130528A field at 
optical/nIR wavelengths. No correction for Galactic extinction 
is applied.} 
\label{table:Optlog}
\centering
\begin{tabular}{lccccc}
\hline\hline
Start Time (UT) & T-T$_{0}$ & Telescope/ & Filter & Exposure time & Magnitude \\
(mid exposure) & (mid days) & Instrument & Grism & (seconds) & (AB) \\
\hline
May 28, 2013, 17:52:42.438 & 0.050 & BOOTES-4/MET & clear & 120 & \textgreater 19.7 (3$\sigma$)\\
May 28, 2013, 17:54:46.385 & 0.052 & BOOTES-4/MET & Sloan r & 180 & \textgreater 19.8 (3$\sigma$)\\
May 28, 2013, 17:57:50.198 & 0.054 & BOOTES-4/MET & clear & 120 & \textgreater 19.7 (3$\sigma$)\\
May 29, 2013, 20:41:42.640 & 1.167 & 1.5m OSN & $I$ & 600 & \textgreater 23.0 (3$\sigma$)\\
\hline
Jun 1, 2013, 03:04:50.211 & 3.409 & 10.4m GTC & Sloan i & 100 $\times$ 9 & 22.87 $\pm$ 0.08\\ 
Jun 1, 2013, 02:41:32.625 & 3.417 & 10.4m GTC & Sloan r & 60 $\times$ 3 & 23.28 $\pm$ 0.01\\
Jun 2, 2013, 02:56:40.228 & 4.427 & 10.4m GTC & Sloan i & 100 $\times$ 1 & 22.74 $\pm$ 0.15\\
Dec 23, 2013, 05:45:16.011 & 208.554 & 4.2m WHT & $J$ & 9 $\times$ 197 & 21.63 $\pm$ 0.35\\
Dec 23, 2013, 06:59:02.247 & 208.610 & 4.2m WHT & $K_{S}$ & 3 $\times$ 432 & 21.43 $\pm$ 0.25\\
Dec 31, 2013, 21:57:12.000 & 217.219 & 6m BTA & B & 300 $\times$ 8 & 23.41 $\pm$ 0.10\\
Apr 05, 2014, 22:37:42.667 & 312.247 & 2.0m LT & Sloan z & 300 $\times$ 36 & 22.41 $\pm$ 0.15\\
\hline
\hline
\end{tabular}
\end{table*} 

\subsection{Millimetre observations}
\label{MilObs}
\indent Millimetre observations were obtained between $\sim$ 1.33 d and $\sim$ 3 d 
after the GRB onset on May 29, 2013 and Jun 02, 2013, via our on-going ToO 
program at PdBI in the French Alps \citep{Guilloteau:1992aa} 
with a compact five antenna D configuration. Observations were conducted 
at 86.7 GHz with a beam size of 5.95" $\times$ 4.81". Data reduction 
was carried out using the GILDAS\footnote{http://www.iram.fr/IRAMFR/GILDAS} software.\
\subsection{Host galaxy spectroscopy}
\label{OptSpe}
\indent Optical spectroscopy with OSIRIS at the 10.4m GTC started on Jun 02, 2013, which is
 $\sim$ 3.4 d after the trigger, using the R1000B (3 $\times$ 600 s exposures) and R500R grism 
(1 $\times$ 600 s exposures). 
The 1.2" slit was positioned on the location of the host galaxy, and 
a 2 $\times$ 2 binning mode was used for data acquisition. The obtained spectra 
were reduced and calibrated following standard procedures using custom tools 
based in IRAF and Python. The standard spectrophotometric stars, Feige92 and 
GD140, were used for flux calibration (for R1000B and R500R, respectively, with a 
2.5"slit). We corrected flux of slit losses for all the spectra using 
the photometry of the corresponding acquisition images.\

\begin{table*}
\caption{mm afterglow flux densities measured at the Plateau de Bure Interferometer.} 
\label{table:radlog}
\begin{tabular}{llccc}
\hline
Start Time (UT) & End Time (UT) & Configuration & Flux density [mJy] & Frequency [GHz] \\
\hline\hline
May 29, 2013, 23:33 & May 30, 2013, 01:42 & 5Dq & 0.38 $\pm$ 0.10 & 86.7 \\
Jun 02, 2013, 13:13 & Jun 02, 2013, 15:03 & 5Dq & $\textless$ 0.07 $\pm$ 0.09 & 86.7 \\
\hline
\end{tabular}
\end{table*}

\begin{figure}
\centering
\includegraphics[width=8cm]{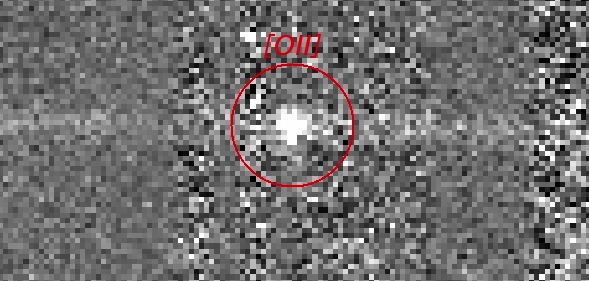}
\caption{The proposed [OII] 3727 emission line in the 2D spectrum of the host galaxy for GRB\,130528A. 
The emission line lies at 8500\AA, implying a redshift 1.250. The 1D spectrum cannot be properly extracted due to its faintness. We determined 
the line flux using photometry, 48 $\pm$ 7.0 $\times$ $10^{-18}$ ergs/s/cm$^{2}$.}
\label{fig:OII}
\end{figure} 

\section{Results and discussions}
\label{Results}
\subsection{No optical/nIR afterglow detections and host galaxy observations at the optical/nIR wavelengths}
\indent No plausible optical/nIR transient was detected down to 19.7 mag (T$_{0}$+1.1 h, clear) at BOOTES-4/MET 
and 23.0 mag (T$_{0}$+1.17 d, $I$-band) at 1.5m OSN telescopes. The potential host galaxy was first revealed using our on-going ToO program at 10.4m GTC in the Sloan r and 
i-bands and is shown in Fig. \ref{fig:fc}. It has 22.9 mag in the 
i-band and 23.3 mag in the r-band. The z-band brightness is revealed with 22.4 mag by LT. The nIR data was reduced under the 
IRAF routine and resulted in magnitudes of 21.6 mag and 21.4 mag in $J$ and $K_{S}$, respectively. The 
BTA B-band observation gave a magnitude of 23.4 mag. All magnitudes are given in Table \ref{table:Optlog} and are presented in the AB system 
(vega to AB offset is following \citealt{Fukugita:1995aa}).\

\subsection{Afterglow detection at mm}
\indent From the mm observations between $\sim$ 1.33 d and $\sim$ 3 d after the GRB, 
we clearly confirm a mm afterglow with a signal-to-noise ratio of 
$\sim$ 4 (at 86.7 GHz) at the position of the putative host galaxy, 
thus confirming the association. The mm source was not detected in the second dataset, 
implying a significant fading of the afterglow flux between the two observations. 
The phase center coordinates are RA(J2000)=$09^{h}$$18^{m}$$02.22^{s}$ and 
Dec(J2000)=+87\textdegree18'05.7" in the outskirts of the XRT error circle 
\citep{DElia:2013aa}. The position coincides well with the astrometrically corrected 
X-ray position \citep{Goad:2013aa}, which enabled us to propose the identification of the host 
galaxy of GRB\,130528A.\

\begin{figure}
\centering
\includegraphics[width=9cm]{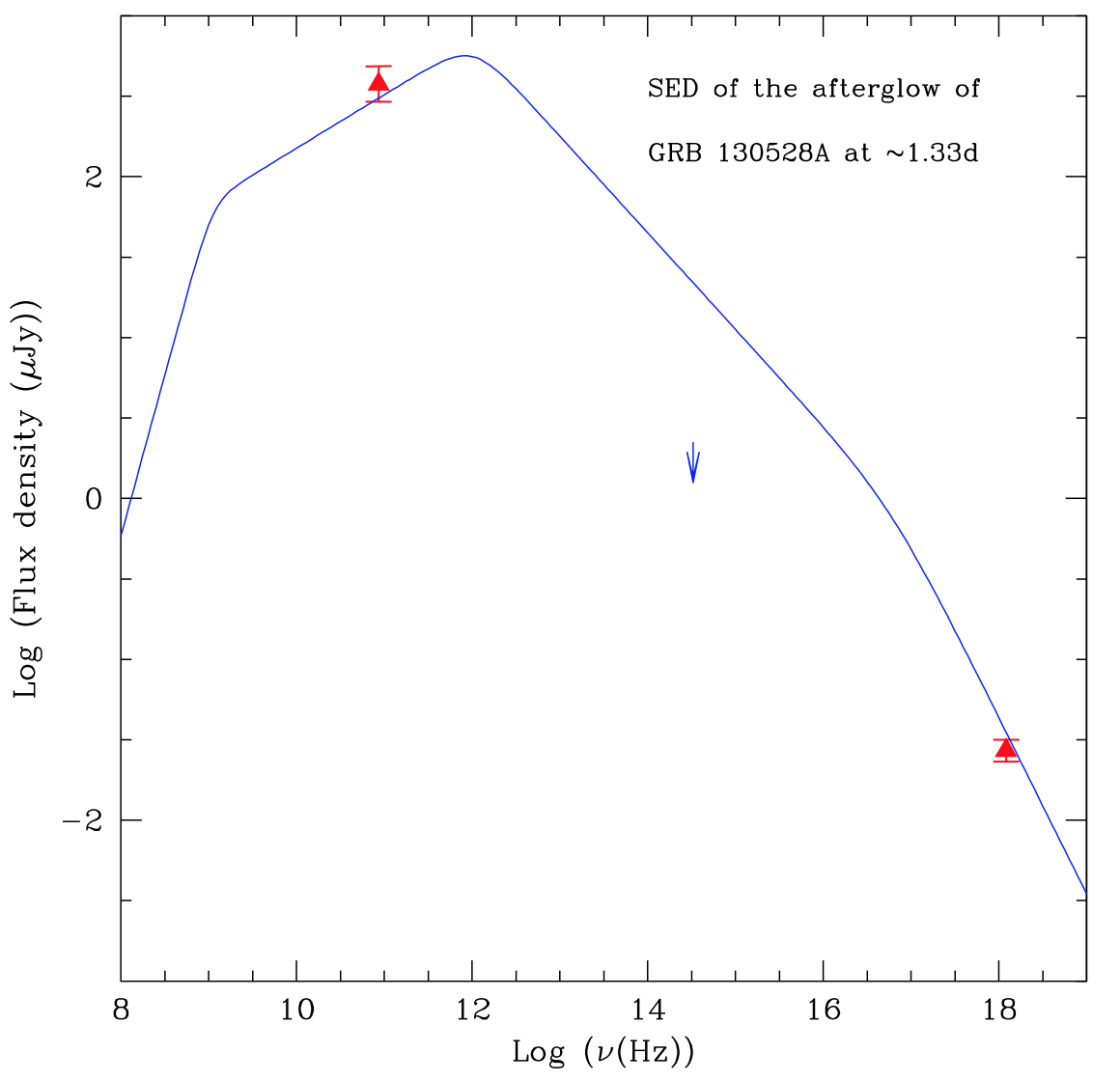}
\caption{The SED of GRB\,130528A afterglow at 1.33 days. A model is 
overplotted with the observed data, assuming 
$\nu_{a}$=1.1 $\times$ $10^{9}$ Hz, $\nu_{m}$=1.1 $\times$ $10^{12}$, 
$\nu_{c}$=6 $\times$ $10^{16}$, F$_{\nu_{max}}$=700 $\mu$Jy, p=2.2, 
and a smoothing parameter s=3 in the case of the ISM model $\nu_{a}$ $\textless$ 
$\nu_{m}$ $\textless$ $\nu_{c}$. The red arrows represent observed flux in the 
radio and X-ray at 1.33 days. The blue arrow shows the $I$-band upper 
limit observed by 1.5m OSN at a similar epoch. The low upper limit can be explained 
by a significant extinction in the line-of-sight.}
\label{fig:sedaft}
\end{figure}  

\subsection{Predicted optical brightness from the afterglow SED}
\label{afSED}
\indent We determined the expected brightness for the optical 
afterglow by following the standard fireball model \citep{Sari:1998}. At 
the time of the radio observation by the PdBI, (i.e., T$_{0}$+1.33 d), 
we took the X-ray flux at a similar epoch from the XRT light curve. The 
over plot of the afterglow SED model (see Fig. \ref{fig:sedaft}) was constructed assuming the following 
parameters: $\nu_{a}$=1.1 $\times$ $10^{9}$ Hz, $\nu_{m}$=1.1 $\times$ $10^{12}$ Hz, $\nu_{c}$=6 $\times$ $10^{16}$ Hz, 
$F_{\nu_{max}}$=700 $\mu$Jy, p=2.2, and a smoothing parameter s=3 in the slow cooling regime \citep{Sari:1999aa}, showing the best overlap to the observed mm and X-ray data. 
We also constrained the parameters of the blast wave as E$_{52}$=10.9, n=0.004 cm$^{-3}$, $\epsilon_{e}$=0.02, and $\epsilon_{b}$=0.01 
by the assumed model parameters using equations 27-30 of \citet{Wijers:1999aa}. We checked the appropriation of p using the 
time sliced XRT spectrum tool\footnote{http://www.swift.ac.uk/xrt$\mathunderscore$spectra} on the timescale that displays less spectral evolution, such as T$_{0}$+22091 - T$_{0}$+115883 s. 
The fit shows the photon index to be $2.02^{+0.44}_{-0.40}$, resulting in a large uncertainty in p. Therefore, we used a universal value of p=2.2 for the energy distribution of the electrons. 
The XRT spectral tool also gave a intrinsic 
value for excess absorption to be $N_{H}$=3.7$^{+2.6}_{-2.0}$ $\times$ $10^{21}$ cm$^{-2}$. 
The modelled SED can be used to estimate the amount of extinction by dust in the 
line-of-sight. The predicted magnitude in $I$-band at 
T$_{0}$=1.33 days is $\sim$ 20.4 mag (vega), and the upper limit produced by 1.5m 
OSN was 22.5 mag (vega). Therefore, it implies a minimum extinction 
A$_{I,min}$$\sim$0.8 mag (A$_{V,min}$$\sim$0.9) at rest frame (after 
galactic extinction correction by \citealt{Schlafly:2011aa}). 
Another independent measurement of the expected UV/optical extinction can be obtained from 
the X-ray absorption to dust-extinction ratio, $N_{H, X}$/$A_{V}$, following \citet{Schady:2010aa}. 
The X-ray absorption $N_{H, X}$ of GRB 130528A with rest frame, which is produced using the “zTBabs” model within Xspec \citep{P.Evans:2009aa}, 
is $N_{H, X}$=2.79 $\pm$ 2.61 $\times$ $10^{22}$ cm$^{-2}$. Using the mean values of 
$N_{H, X}$/$A_{V}$=3.3 $\times$ $10^{22}$, 3.4 $\times$ $10^{22}$ and 2.1 $\times$ $10^{22}$ cm$^{-2}$ for the different extinction models 
(Small Magellanic Cloud (SMC), Large Magellanic Cloud (LMC), and Milky Way (MK), see \citealt{Schady:2010aa}), 
$A_{V}$ was found to be 0.84, 0.82, and 1.32 mag for SMC, LMC, 
and MW, respectively. These values are consistent with the previous findings from the afterglow SED.
 
\begin{figure}
\centering
\includegraphics[width=9cm]{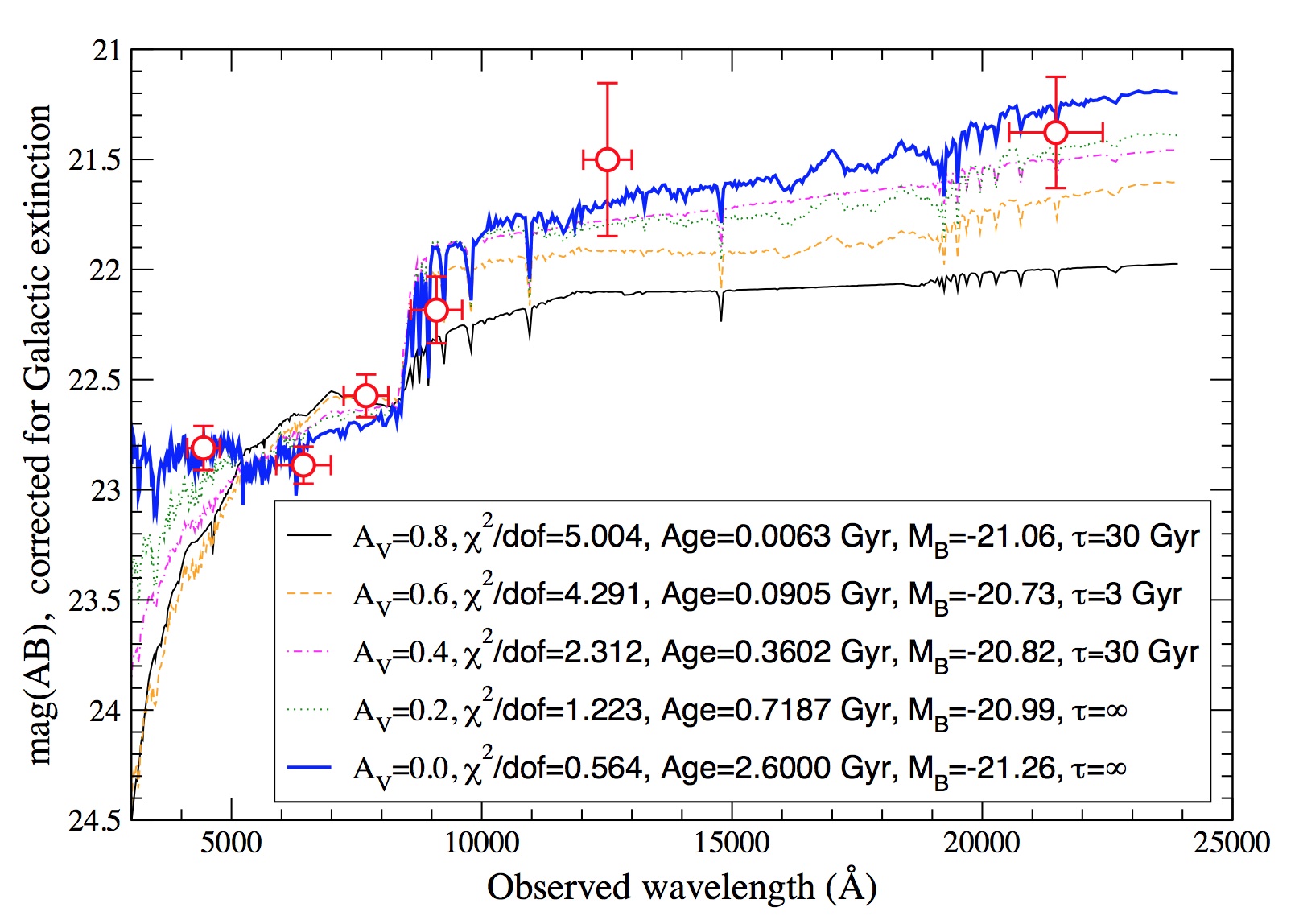}
\caption{The SED of GRB\,130528A host galaxy in the B,r,i,z,$J$, and $K_{S}$ bands. The AB magnitudes are 
corrected considering for the Galactic extinction E(B-V)=0.144 \citep{Schlafly:2011aa}. 
The thick line shows the best fit ($\chi^{2}$/$d.o.f.$=0.564) achieved with $A_{V}$=0, 
a stellar population with an age of 2.6 Gyr, M$_{B}$=-21.16, and Solar metallicity. 
The rest of the lines show the evolution of the SED fit when $A_{V}$ increases gradually 
from 0 to 0.8. As seen, the fit gets worse when $A_{V}$ grows.}
\label{fig:sed}
\end{figure} 

\subsection{Redshift determination and the star-formation rate}
\label{redshifts}
\indent The single emission line at 8500$\AA$ in the GTC spectra likely corresponds with [OII] 3727$\AA$ at a redshift 
$z$=1.250 $\pm$ 0.001, however, due to the low SNR of the spectrum, the 1D 
projection of the optical spectrum could not be extracted. The [OII] 
line in 2D spectrum is shown in Fig. \ref{fig:OII}. The emission line 
flux is 48 $\pm$ 7.0 $\times$ $10^{-18}$ ergs/s/cm$^{2}$, as measured by 
line photometry. Using this value, we can derive a lower limit on the SFR by applying the calibration of \citet{Kennicutt:1998aa}, which is SFR($M_{\odot}$/yr)=1.4 $\pm$ 
0.4 $\times$ $10^{41} $ $L_{\rm [O II]}$. At a redshift $z$=1.25, the 
corresponding luminosity distance is $D_{\rm L}$=8.768 Gpc (assuming 
$H_{0}$=71 km s$^{-1}$Mpc$^{-1}$, $\Omega_{\rm m}$=0.27, and 
$\Omega_{\Lambda}$=0.73). Taking the lower limit for 
the optical extinction into account, this implies SFR($M_{\odot}$/yr) > 6.18 
$M_{\odot}$/yr. It signifies a SSFR$\sim$5.3 $M_{\odot}$/yr $(L/L^{\star})^{-1}$ (assuming 
$M^{\star}_{B}=-21$), which is a low value compared to other SSFRs estimated in long GRB host galaxies \citep{Christensen:2004aa}.

\subsection{The host galaxy spectral energy distribution}
\indent Broadband observations were matched using the Hyperz code 
\citep{Bolzonella:2000aa} to synthetic SED templates based on the 
GISSEL 98 library \citep{Bruzual-A.:1993aa}. The time evolution of the 
SFR for each template is represented by an exponential model, which is 
SFR $\propto$ $\exp(-t/\tau)$, where $\tau$ is the SFR timescale. Eight values of $\tau$ were explored (0,1,2,3,5,15,30,and $\infty$ 
Gyr) and the initial mass function (IMF) given by 
\citet{Miller:1979aa} was assumed. The impact of the metallicity 
(which is expected to be minor, as tested by \citealt{Bolzonella:2000aa})
was considered, using both solar metallicity and evolving 
metallicity templates, which assumed an instantaneous recycling of 
heavy elements. For each $\tau$ value, the extinction ($A_{V}$) and 
the dominant stellar population age were left as free parameters. The 
host galaxy magnitudes were corrected by Galactic reddening by assuming an 
E(B-V)=0.144 \citep{Schlafly:2011aa}.\\ 
\indent To check the stability of our SED solution, we first let the 
redshift vary from 0.1 to 5. The photometric redshift yields 
$z$=$1.397^{+0.097}_{-0.085}$ (68\% confidence interval) are fairly 
consistent with the spectroscopic redshift. Thus, the redshift of the SED 
fit was fixed to the spectroscopic redshift value (see 
Sect. \ref{redshifts}). As displayed in Fig. \ref{fig:sed}, the best 
fit is obtained with a Solar metallicity and intrinsically bright 
(M$_{B}$=-21.16) and low extinction ($A_{V}$$\sim$0) galaxy, which dominated 
by an evolved (age$\sim$2.6 Gyr, $\tau$=$\infty$) stellar 
population. If the metallicity is assumed to be evolving, the derived 
values are qualitatively equivalent ($A_{V}$$\sim$0, Age$\sim$1.434 
Gyr, $\tau$=5 Gyr, M$_{B}$=-21.03, $\chi^{2}$/$d.o.f.$=0.754). Due to 
the lack of other emission lines within our spectral window, we cannot 
constrain the metallicity of the host galaxy.\\
\indent This is not the first case where a low (global) extinction 
galaxy harbours a highly extinguished afterglow (for instance 
GRB\,000210 or GRB\,000418; \citealt{Gorosabel:2003aa, Gorosabel:2003b}). 
This apparent contradiction might be explained by 
line-of-sight dust that is probably close to the progenitor birth place. 
However, we have to be cautious about the uniqueness of the SED 
solution. As a sanity check of a possible age-extinction degeneracy, 
we studied the SED fit evolution when the host $A_{V}$ ranged from 
$A_{V}$=0 to $A_{V}$=0.8 (see Fig. \ref{fig:sed}). As expected, the 
dominant stellar population age decreases when $A_{V}$ grows. 
However, the impoverishment of the fit is clear when $A_{V}$ increases; 
$A_{V}$ values larger than 0.8 (not shown in the plot for visual clarity) 
provide an even worse fit. We, thus, conclude that the $A_{V}$=0 SED fit is a solid solution.\\ 
\indent Unfortunately for the redshift of the host, it is impossible that the 
$H_{\alpha}$ and $H_{\beta}$ emission lines do fall in the 
$JHK$-band atmospheric windows, so no Balmer decrement 
(and hence direct host extinction spectroscopic check) 
measurement would be possible from the ground.\\
\indent The derived high $\tau$ and old stellar population age 
are consistent with low SSFR (see Sect. \ref{redshifts} in this paper), 
therefore a feasible picture could be one in which the main episode(s) 
of star-formation of this host remains constant in time or slowly decays in intensity. 
The host of GRB\,130528A also has a red colour, (R-K)$_{AB}$=1.54 
(Sloan r-filter is transformed to the R using Lupton 2005\footnote{http://classic.sdss.org/dr5/algorithms/sdssUBVRITransform.html}), 
this colour excess yields a similar characteristic to the hosts of the dustiest afterglows, as shown in \citet{Kruhler:2011aa}.
 
\section{Conclusions}
\label{Conclusions}
\indent
In this work, we have shown that the darkness of the long duration "dark" 
GRB\,130528A was likely due to the high absorption close to the GRB 
birth place, lying in a galaxy at $z$=1.25, which is pinpointed thanks to the mm 
detection by PdBI. Based on optical/nIR host galaxy observations 
at GTC, WHT, BTA, and LT, we infer that the GRB\,130528A occurred in a 
low extinction ($A_{V}$$\sim$0), aged (dominate stellar population of 
2.6 Gyr), red ((R-K)$_{AB}$=1.54), and luminous (M$_{B}$=$-$21.16) host. 
However, this host galaxy seems different with respect to the main body of long-GRB hosts \citep{Christensen:2004aa} 
and is consistent with the characteristics of the dustiest hosts shown 
by the latest statistical studies \citep{Perley:2013aa, Rossi:2012aa}. 
Through mm and nIR observations by PdBI and 1.5m OSN at similar 
epochs, we infer the relative extinction along the line-of-sight towards the GRB\,130528A, which is 
$A^{GRB}_{V}$$\geq$ 0.9 mag in rest frame. We also show that this result is consistent 
with the expected UV/optical extinction from rest frame $N_{H, X}$. 
The inconsistency between the significant 
extinction expected from the afterglow SED model and low external 
extinction determined from the host galaxy SED could be reconciled if 
the GRB was located in a high density environment, such as a local 
molecular cloud. 

\begin{acknowledgements}

This work is partly based on observations carried out with the 0.6m BOOTES-4/MET 
in China, with the Gran Telescopio Canarias (GTC), installed in the 
Spanish Observatorio del Roque de los Muchachos of the Instituto de 
Astrofísica de Canarias, in the island of La Palma and with the IRAM 
Plateau de Bure Interferometer. IRAM is supported by INSU/CNRS 
(France), MPG (Germany) and IGN (Spain). The Liverpool Telescope is operated on the island 
of La Palma by Liverpool John Moores University in the Spanish Observatorio del Roque de 
los Muchachos of the Instituto de Astrofisica de Canarias 
with financial support from the UK Science and Technology Facilities Council. 
Alexander Moskvitin supported 
by Research Program OFN-17 of the Division of Physics, Russian Academy 
of Sciences. We acknowledge the support of F. J. Aceituno (OSN 
observatory) and the Spain’s Ministerio de Ciencia y Tecnología 
through Projects AYA2009-14000-C03-01/ESP, AYA2011-29517-C03-01, 
AYA2012-39362-C02-02, and Creative Research Initiatives of NRF in 
Korea (Research Center of MEMS Space Telescope). This work made use of 
data supplied by the UK {\it Swift} Science Data Centre at the 
University of Leicester. We acknowledge the use of public data from the {\it Swift} data archives and the service 
provide by the gamma-ray burst Coordinates Network (GCN) and BACODINE 
system, maintained by S. D. Barthelmy. We thank the referee for carefully reading this manuscript and providing 
valuable comments, which helped improve the text substantially. We also acknowledge valuable discussions and comments from 
Z. Lucas Uhm.
\end{acknowledgements}

\bibliographystyle{aa}
\bibliography{130528a2}
\end{document}